\documentclass[12pt,reqno]{amsart}

\usepackage{amssymb,amsmath,amsthm,mathtools,wasysym,calc,verbatim,enumitem,tikz,hyperref,url,mathrsfs,cite,fullpage,bbm,comment} 
\mathtoolsset{showonlyrefs,showmanualtags}
\usepackage{scalerel}
\usepackage{stackengine}
\stackMath

\newtheorem{theorem}{Theorem}

\newtheorem{lemma}[theorem]{Lemma}
\newtheorem{claim}[theorem]{Claim}
\newtheorem{corollary}[theorem]{Corollary}

\theoremstyle{remark}
\newtheorem*{remark*}{Remark}

\numberwithin{theorem}{section}

\renewcommand{\phi}{\varphi}

\renewcommand{\leq}{\le}
\renewcommand{\geq}{\ge}

\newcommand{\cE}{\mathcal E}

\newcommand{\cA}{\mathcal A}
\newcommand{\cB}{\mathcal B}

\newcommand{\E}{\mathbb{E}}

\def\1{\mathbbm{1}}

\renewcommand{\le}{\leqslant}
\renewcommand{\ge}{\geqslant}

\DeclareMathOperator{\clique}{CLIQUE}
\DeclareMathOperator{\con}{AND}
\DeclareMathOperator{\dis}{OR}
\DeclareMathOperator{\nt}{NOT}
\DeclareMathOperator{\size}{Size}
\DeclareMathOperator{\monsize}{MonSize}

\usepackage{hyperref}
\usepackage{cleveref}

\crefname{theorem}{theorem}{theorems}
\Crefname{theorem}{Theorem}{Theorems}
\crefname{lemma}{lemma}{lemmas}
\Crefname{lemma}{Lemma}{Lemmas}
\crefname{corollary}{corollary}{corollaries}
\Crefname{corollary}{Corollary}{Corollaries}
\crefname{conjecture}{conjecture}{conjectures}
\Crefname{conjecture}{Conjecture}{Conjectures}
\crefname{construction}{construction}{constructions}
\Crefname{construction}{Construction}{Constructions}
\crefname{observation}{observation}{observations}
\Crefname{observation}{Observation}{Observations}
\crefname{proposition}{proposition}{propositions}
\Crefname{proposition}{Proposition}{Propositions}
\crefname{claim}{claim}{claims}
\Crefname{claim}{Claim}{Claims}
\crefname{problem}{problem}{problems}
\Crefname{problem}{Problem}{Problems}
\crefname{fact}{fact}{facts}
\Crefname{fact}{Fact}{Facts}
\crefname{definition}{definition}{definitions}
\Crefname{definition}{Definition}{Definitions}
\crefname{process}{process}{processes}
\Crefname{process}{Process}{Processes}
\crefname{question}{question}{questions}
\Crefname{question}{Question}{Questions}
\crefname{remark}{remark}{remarks}
\Crefname{remark}{Remark}{Remarks}
\crefname{example}{example}{examples}
\Crefname{example}{Example}{Examples}
\crefname{appendix}{appendix}{appendices}
\Crefname{appendix}{Appendix}{Appendices}
\crefname{subsection}{subsection}{subsections}
\Crefname{subsection}{Subsection}{Subsections}
\crefformat{enumi}{#2item~#1#3}
\Crefformat{enumi}{#2Item~#1#3}
\crefmultiformat{enumi}{items~#2#1#3}{ and~#2#1#3}{, #2#1#3}{, and~#2#1#3}
\crefrangeformat{enumi}{items~#3#1#4--#5#2#6}

\begin{document}

\title{Circuit complexity of $\clique$\\ with a logarithmic negation budget} 

\author{Bruno Andrades}

\address{IMPA, Estrada Dona Castorina 110, Jardim Bot\^anico, Rio de Janeiro, 22460-320, Brasil}\email{bruno.andrades@impa.br}

\thanks{The author was supported by CNPq.}

\begin{abstract}
We show that any polynomial-size circuit computing $\clique(n,\lfloor n/2\rfloor)$ uses at least $\frac12\log n-\log\log n-O(1)$ $\nt$ gates. This is optimal up to a constant factor by a result of Fischer.
\end{abstract}

\maketitle

\vspace{-2em}
\section{Introduction}
Proving super-polynomial lower bounds for the circuit complexity of Boolean functions is one of the main problems studied in circuit complexity. This is best illustrated by its consequence on central problems, like $P\overset{?}{=}NP$, where showing super-polynomial circuit complexity bounds on certain natural functions would be enough to show that $P\neq NP$. In contrast, we do have machinery to deal with the related problem when we restrict ourselves to monotone circuits. That is, circuits without $\nt$ gates; in a seminal paper, Razborov \cite{R85} showed that if we denote by $\monsize f$ the smallest size of a monotone circuit computing $f$, then\footnote{We let $\log$ denote the binary logarithm while $\ln$ denotes the logarithm in base $e$.}
$$\monsize\clique\left(n,\left\lfloor\frac14\ln n\right\rfloor\right)=n^{\Omega(\log n)}$$
where $\clique(n,k):\{0,1\}^{\binom n2}\to\{0,1\}$ is the function that outputs $1$ if and only if its input, when seen as an $n$-vertex graph in the natural way, contains a clique of size $k$. Thus proving for the first time super-polynomial lower bounds for the monotone circuit complexity of a function known to be in $NP$. We will henceforth restrict our attention to that of circuit lower bounds for the function $\clique$.

While the monotone circuit size question has received a lot of attention (e.g. \cite{AB85,BM25,R26}), monotone circuit lower bounds do not necessarily translate into lower bounds for the general circuit lower bound problem. A natural interpolation between the monotone problem and the general one is to consider a negation-budget version. In particular, for $t\geq0$ an integer define $\size_tf$ as the smallest size of a circuit that computes $f$ and contains at most $t$ $\nt$ gates. In \cite{F75}, Fischer shows that the size of a circuit computing $f:\{0,1\}^N\to \{0,1\}$ and that of a circuit that computes $f$ with at most $\lceil\log(N+1)\rceil$ negations are polynomially related. Thus, if for $t=\left\lceil\log\left(\binom n2+1\right)\right\rceil\sim 2\log n$ one could show that there exists $k=k(n)$ such that
$$\size_t\clique(n,k)=n^{\omega(1)}$$
one could conclude that $P\neq NP$. In this direction, Amano and Maruoka \cite{AM05} show that for $t=\frac16\log\log n$ one has
$$\size_t\clique\left(n,(\log n)^{3\sqrt{\log n}}\right)=n^{\omega(1)}.$$
We will improve this, and get the optimal order of magnitude for $t$, by showing the following theorem

\begin{theorem}\label{thm:main} Any circuit computing $\clique(n,\lfloor n/2\rfloor)$ using at most $\frac12\log n-\log\log n-\omega(1)$ negations has super-polynomial size.
\end{theorem}
This indeed is tight up to a factor of $4+o(1)$, as proving the same result for $t=\left\lceil\log\left(\binom n2+1\right)\right\rceil$ would imply the same superpolynomial lower bound in the unrestricted setting. The proof combines the approach of Amano and Maruoka \cite{AM05} with a recent result of Rao \cite{R26}, and also replaces their iterative pigeonhole approach by a more efficient telescoping argument. We give a detailed outline of the proof in \Cref{sec:overview}.

\section*{AI declaration}
The proof was found with no help from any LLM outside of literature research. After the author had come up with the proof of \Cref{thm:main}, an LLM was prompted to improve the results (without having access to the author's proof) and it managed to improve the result to the order $t\geq (1-o(1))\log n$. Very recently, after the author prompted an LLM with both proofs, it produced a bound of the form $t \geq (2 - o(1)) \log n$ by combining ideas from both. We plan to provide an exposition of this proof in the near future.
\section*{A remark on an independent result}
After the author had completed his proof, he was made aware of the result of of Karde\c s \cite{K26} that deals with our same problem (and also some related questions). In particular, they get an analogous result to \Cref{thm:main_contrapositive} but with a threshold of the form $(1-o(1))\log n$, better than our result, but worse than the $(2-o(1))\log n$ obtained with the help of AI.
\section{Overview of the proof}\label{sec:overview}
The first step of the proof will be to apply \cite[Theorem 3.2]{AM05}, which roughly states that for any monotone Boolean function $f:\{0,1\}^N\to\{0,1\}$ and integer $t\geq 0$, one may find ``few" monotone circuits $f_i$ such that $\monsize f_i\leq \size_t f$ and that every pair of graphs that are distinguished\footnote{Given two disjoint subsets $\cA,\cB\subseteq \{0,1\}^N$, we say that $f$ distinguishes them if $f(A)\neq f(B)$ for every $A\in \cA$ and $B\in \cB$. For our applications it will hopefully be clear which subset should get the value $0$ and which one should get the value $1$.} by $f$, one of the $f_j$ also distinguishes. Formally, define
$$\partial f:=\left\{(x,y)\in \{0,1\}^N\times\{0,1\}^N:||x-y||_1=1,\;f(x)=0,\;f(y)=1\right\}$$
to be the (ordered) edge boundary of $f$. The following theorem was proved by Amano and Maruoka \cite{AM05}
\begin{theorem}[Theorem 3.2 in \cite{AM05}]\label{thm:boundary_cover}
	Let $N\geq1$ and $t\geq 0$ be integers, and let $f:\{0,1\}^N\to\{0,1\}$ be a monotone function. Then, there are $\alpha=2^{t+1}-1$ not necessarily distinct monotone functions $f_1,\ldots, f_\alpha:\{0,1\}^N\to\{0,1\}$ such that
    $$\partial f\subseteq\bigcup_{j=1}^\alpha \partial f_j$$
    and $\monsize f_i\leq \size_t f$ for every $i\in[\alpha]$.
\end{theorem}
We will focus on a particular subset of $\partial\clique(n,k)$, namely, the elements arising from balanced complete $(k-1)$-partite graphs. Indeed, if $Y$ is a balanced complete $(k-1)$-partite graph, and $Y^+$ is obtained by adding one edge to one of the non-singleton classes of $Y$, whenever such a class exists, then
$$(Y,Y^+)\in\partial\clique(n,k).$$

In particular, we'll use the following immediate consequence of \Cref{thm:boundary_cover} 
\begin{corollary}\label{coro:boundary_cover}
    Let $n\geq k\geq 2$ and $t\geq0$ be integers. Then, there exists $\alpha=2^{t+1}-1$ monotone functions $f_1,\ldots, f_\alpha:\{0,1\}^{\binom{n}{2}}\to\{0,1\}$ such that for every balanced complete $(k-1)$-partite graph $Y$ one has
    $$\sum_{j=1}^\alpha \mathbf1[f_j(Y)=0,f_j(Y^+)=1]\geq 1$$
    and $\monsize f_j\leq\size_t\clique(n,k)$ for every $j\in[\alpha]$.
\end{corollary}
Going forwards, whenever we say Tur\'an graph, we will be referring to a balanced complete $(k-1)$-partite graph, while the number of vertices will be made explicit every time. Whenever we have $L$ vertices in such a graph, it will be called a $T_{k-1}(L)$. 

The main idea, which is also present in \cite{AM05}, is that whenever we pick a random $(Y,Y^+)$, since it's deterministically in $\partial\clique(n,k)$, there must be some $f_j$ like in \Cref{thm:boundary_cover} that covers it, and therefore, if $t$ is small, a pigeonhole argument shows that some of the $f_j$ must cover a decent proportion of the pairs $(Y,Y^+)$, and can therefore, by monotonicity, distinguish a decent proportion of Tur\'an graphs from large cliques. In \cite{AM05} they deal with this by proving that if a monotone circuit accepts a decent proportion of cliques of a specific size, and rejects a decent proportion of Tur\'an graphs of a different specific size, then, the circuit must have large size\footnote{For a more detailed statement, see {Theorem 4.1} in \cite{AM05}.}, they then apply this to many different sizes of Tur\'an graphs and cliques. An iterative pigeonhole argument shows that if all the $f_i$ have small size, there must be some pair $(Y,Y^+)$ that is not covered by any of the $f_j$. 

We deal with this same phenomenon in a somewhat similar manner, though we avoid the need for an iterative pigeonhole argument by slightly changing the definition of what it means to be good at distinguishing cliques from Tur\'an graphs, and instead replace it with a more efficient telescoping argument (see \Cref{claim:telescoping}). In particular, \Cref{claim:telescoping} tells us that if for some integers $k\leq L_0<\dots<L_m\leq n$ a monotone function $g$ struggles to distinguish cliques of size $L_i$ from Tur\'an graphs of size $L_{i+1}$, in the sense that $\E g(X_{L_i})-\E g(Y_{L_{i+1}})\leq m^{-1}$, where $X_{L}$ and $Y_L$ are uniformly random cliques and Tur\'an graphs on $L$ vertices respectively, then $g$ distinguishes a small proportion of the pairs $(Y_{L_i},Y_{L_i}^+)$. Therefore, if we manage to show that the $f_j$ are bad at distinguishing cliques from Tur\'an graphs, then we can show that we need many $f_j$ and therefore we need a large value of $t$, this is \Cref{lem:size_bound}. Replacing this iterative pigeonhole argument by \Cref{lem:size_bound} provides our main saving over \cite{AM05}.

Here enters the second difference with the approach of \cite{AM05}. We dispose of their \cite[Theorem 4.1]{AM05}, by instead lower-bounding the size of an $f_j$ that's good at distinguishing cliques from Tur\'an graphs via some reductions. In particular, we will prove that if a function $g$ is good at distinguishing Tur\'an graphs from cliques, in the sense that $\E g(X_{L_i})-\E g(Y_{L_{i+1}})\geq m^{-1}$, then, by choosing $d=L_{i+1}-L_i$ carefully, for a suitable value of $r$, one can use this $g$ to construct a slightly larger monotone circuit on $K_{r,r}$ that distinguishes graphs with perfect matchings to those with no large matchings. We will then be in a setting where we'll be able to apply the following result of Rao \cite{R26}.
\begin{theorem}[Theorem 1 in \cite{R26}]\label{thm:rao} For $q<r\leq N/2$, any monotone circuit that can distinguish $N$-vertex bipartite graphs with matchings of size $r$ from graphs that do not have a matching of size $q$ must have monotone size at least $2^{\Omega(q/\sqrt r)}$.  
 \end{theorem}

This, by the above discussion, will imply that $g$ must have had large monotone size. The details of this reduction are the main subject of \Cref{sec:approximate_to_exact} and \Cref{sec:reduction}. 

Once we know that any function of small monotone size struggles to distinguish cliques from Tur\'an graphs, we will be able to conclude via \Cref{lem:size_bound} that any circuit calculating $\clique(n,k)$ with polynomial size, must have many negations, as otherwise, one of the $f_j$ guaranteed by \Cref{thm:boundary_cover} must have at the same time small monotone size, and be good at distinguishing cliques from Tur\'an graphs, which by our discussion above, is a contradiction after a careful tuning of the parameters. This will conclude the proof. 

The rest of the paper is organised as follows: In \Cref{sec:preliminaries} we state and prove \Cref{lem:size_bound} which tells us how to get bounds on $t$ from the fact that functions with small monotone size struggle to distinguish cliques from Tur\'an graphs. \Cref{sec:approximate_to_exact} and \Cref{sec:reduction} will be devoted to showing that functions that are good at distinguishing cliques from Tur\'an graphs in the sense that $\E g(X_{L_i})-\E g(Y_{L_{i+1}})\geq m^{-1}$ must have large monotone circuit complexity. Specifically, \Cref{sec:approximate_to_exact} will contain a short argument that shows how the existence of a function satisfying the above condition, implies the existence of an at most polynomially larger monotone circuit on a smaller set of size $L_{i+1}$ that perfectly distinguishes cliques of size $L_{i}$ from copies of $T_{k-1}(L_{i+1})$. Then, in \Cref{sec:reduction} we'll see how any such monotone circuit can be used to construct a monotone circuit of the same size on $K_{r,r}$ that distinguishes graphs with perfect matchings from those with no large matchings, where $r=L_{i+1}-k+1$. Then we will apply \Cref{thm:rao} to conclude that such a circuit must be large. Finally, in \Cref{sec:main_proof}, we will make the parameter choice for $m,k, L_0,\ldots, L_m$ and put all the pieces together to prove \Cref{thm:main_contrapositive}. Ultimately, we will choose 
$$k=\lfloor n/2\rfloor,\qquad m=\Theta_a\left(\frac{\sqrt n}{\log n}\right),\qquad L_i=\frac32k+id$$
where $d=Ca\sqrt n\log n$. \Cref{lem:size_bound} will then give that
$$t\geq\log m-O(1)=\frac12\log n-\log\log n-O_a(1).$$
In \Cref{sec:concluding_remarks} we discuss the optimality of the order in our result, and some of the (rather daunting) hurdles that one would need to overcome if one wanted to fully solve the problem and thereby establish $P\neq NP$.

\section{From cliques and Tur\'an graphs to a negation budget}\label{sec:preliminaries}

In this section we will see how to deduce bounds on the number of negations of a circuit that struggles to distinguish cliques of some size from slightly larger Tur\'an graphs. This will be the job of the following lemma

\begin{lemma}\label{lem:size_bound}
Let $k\leq L_0<L_1<\ldots<L_m\leq n$ be integers and suppose that 
\begin{equation}
\E g(X_{L_i})-\E g(Y_{L_{i+1}})
\leq m^{-1}
\tag{1}
\end{equation}
for every monotone function $g$ with $\monsize g\leq s$ and every $i\in \{0,\ldots, m-1\}$. Then, 
$$\size_t\clique(n,k)\leq s\quad\Longrightarrow\quad t\geq \log m -O(1)$$
\end{lemma}
Recall that if $Y$ is a copy of $T_{k-1}(L)$, then $Y^+$ denotes the graph obtaining by adding an edge to one of its non-singleton classes. The proof will be an application of \Cref{coro:boundary_cover} plus the simple observation that for any monotone function $g$, one has
$$\mathbf{1}\big[g(Y_L)=0,g(Y^+_L)=1\big]\leq g(K_S)-g(Y_L)$$
where $K_S$ denotes a clique on the support of $Y_L$. A simple averaging argument then finishes the proof
\begin{proof}[Proof of \Cref{lem:size_bound}]
Notice that if $\size_t\clique(n,k)\leq s$, then by \Cref{coro:boundary_cover} there must exist $f_1,\ldots, f_\alpha$, each with a monotone circuit of size at most $s$ such that 
$$1\leq\sum_{j=1}^\alpha \mathbf1[f_j(Y_{L_i})=0,f_j(Y_{L_i}^+)=1].$$
Taking expectations and averaging over $i$ one gets
\begin{equation}
    1\leq\sum_{j=1}^\alpha\frac1m\sum_{i=0}^{m-1}\mathbb P(f_j(Y_{L_i})=0,f_j(Y_{L_i}^+)=1).\tag{2}
\end{equation}
To upper bound the right hand side, we will show that under the hypotheses of \Cref{lem:size_bound}, the following claim holds

\begin{claim}\label{claim:telescoping} We have
 $$\frac1m\sum_{i=0}^{m-1}\mathbb P(g(Y_{L_{i}})=0,g(Y_{L_{i}}^+)=1)\leq\frac2m$$
 for every $g$ with $\monsize g\leq s$.
\end{claim}

\begin{proof}[Proof of \Cref{claim:telescoping}]
By monotonicity, for every $L$ one has
$$\mathbf{1}[g(Y_L)=0,g(Y^+_L)=1]\leq g(K_S)-g(Y_L)$$
where $K_S$ is the $n$-vertex graph that consists of a clique on the support of $Y_L$ and $n-L$ isolated vertices. Therefore, taking expectations and noticing that $K_S\sim X_L$ we get that
$$\frac1m\sum_{i=0}^{m-1}\mathbb P(g(Y_{L_{i}})=0,g(Y_{L_{i}}^+)=1)\leq\frac1m\sum_{i=0}^{m-1}\E g(X_{L_i})-\E g(Y_{L_{i}})$$
This right hand side can be rewritten as
\begin{align*}
\frac1m\left(\E g(Y_{L_m})-\E g(Y_{L_{0}})\right)+\frac1m\sum_{i=0}^{m-1}\E g(X_{L_i})-\E g(Y_{L_{i+1}})\leq \frac2m
\end{align*}
where the first term is bounded using that $g$ is a Boolean function and the second one by (1). This proves the claim.
\end{proof}

Applying \Cref{claim:telescoping} to each $f_j$ in (2) we obtain
$$1\leq\sum_{j=1}^\alpha\frac1m\sum_{i=0}^{m-1}\mathbb P(f_j(Y_{L_i})=0,f_j(Y_{L_i}^+)=1)\leq\frac{2}m(2^{t+1}-1)$$
which in turn implies $t\geq \log m-O(1)$ as desired.

\end{proof}
\section{Efficiently distinguishing cliques from Tur\'an graphs}\label{sec:approximate_to_exact}
We will see that if a monotone circuit $g$ of size $s$ satisfies that 
$$\E g(X_{L-d})-\E g(Y_{L})\geq m^{-1}$$
 then one can increase the size of $g$ only polynomially in $m$ and $L$ to obtain a circuit on $K_L$ that distinguishes copies of $K_{L-d}$ from copies of $T_{k-1}(L)$
 \begin{lemma}\label{lem:approximate_to_exact}
     Let $g$ be a monotone circuit of size $s$ on $K_n$ such that for some $m\geq 1$, $L\geq 2$ and $d<L$ one has
     \begin{equation}
         \E g(X_{L-d})-\E g(Y_{L})\geq m^{-1}\tag{3}
     \end{equation}
    then, there exists a monotone circuit $F$ on $K_L$ of size at most $\ell s+\ell\log \ell$ that distinguishes copies of $K_{L-d}$ from copies of $T_{k-1}(L)$, where $\ell=O(m^{2}L\log L)$.
 \end{lemma}
 Before going into the proof, it's perhaps useful to sketch it. The first step of the proof will be to show via a simple averaging argument that if the hypotheses of the lemma hold, then there must exist some set $W$ of size $L$ where $g$ is good at distinguishing copies of $K_{L-d}$ from copies of $T_{k-1}(L)$, both on $W$, in the sense that 
 $$\E g(X_{L-d}^W)-\E g(Y^W_L)\geq m^{-1}.$$
 Where $X_{L-d}^W$ and $Y_L^W$ denote a uniformly random copy of $K_{L-d}$ on the vertex set $W$ and a uniformly random copy of $T_{k-1}(L)$ on the vertex set $W$ respectively. Ultimately, $W$ will be the vertex set of size $L$ where we will find the desired circuit $F$. The idea to construct $F$ will be the following: given a graph $x$ on the vertex set $W$, we will compute $g$ at the image of $x$ under a series of randomly chosen permutations of $W$. By Chernoff, if $x$ is a copy of $K_{L-d}$, the average of these images will be tightly concentrated around $a:=\E g(X_{L-d}^W)$, while if it is a copy of $T_{k-1}(L)$, it will be tightly concentrated around $b:=\E g(Y_{L}^W)$. We will design the $F$ exactly to discern whether this average is closer to $a$ or to $b$. This will be done by choosing some value between $a$ and $b$, and letting $F$ be the indicator of whether this average exceeds this fixed value or not. By the main theorem of \cite{AKS83}, given the values of $g$ at these points, evaluating $F$ requires a circuit of small size. Explicitly, we use the following version of the main theorem of \cite{AKS83}, as stated in \cite{K89}
 \begin{theorem}[Theorem 1 in \cite{K89}]\label{thm:AKS} For every $N\geq 1$ and $\tau\geq0$, let 
 $$f_{N,\tau}(x_1,\ldots, x_N)=\mathbf1\left[\sum_{i=1}^Nx_i\geq\tau\right]$$
then
$$\monsize f_{N,\tau}=O(N\log N)$$
 \end{theorem}
 \begin{proof}[Proof of \Cref{lem:approximate_to_exact}]
    Observe that if $W^*\subseteq[n]$ is a uniformly randomly chosen $L$-set, then $X_{L-d}^{W^*}\sim X_{L-d}$ and $Y_{L}^{W^*}\sim Y_{L}$. It follows that if (3) holds then, there exists some realization $W$ of $W^*$ such that 
    $$\E g(X_{L-d}^{W})-\E g(Y^{W}_{L})\geq m^{-1}$$
    Let $a=\E g(X_{L-d}^{W})$ and $b=\E g(Y_{L}^{W})$. Choose a sufficiently large constant $C$ and set $\ell= Cm^{2}L\log L$. Pick
    $$\tau\in\left(\left(b+\frac1{3m}\right)\ell,\left(a-\frac1{3m}\right)\ell\right)$$
     arbitrarily. We will show that there exists some collection $\pi_1,\ldots,\pi_\ell$ of permutations of $W$ such that the function
     $$F(x)=\mathbf 1\left[\sum_{i=1}^\ell g(\pi_ix)\geq \tau\right]$$
     distinguishes copies of $K_{L-d}$ from copies of $T_{k-1}(L)$. In other words, whenever $x$ is a copy of $K_{L-d}$ and $y$ is a copy of $T_{k-1}(L)$
     \begin{equation}
         F(x)=1\qquad \text{and}\qquad F(y)=0.\tag{4}
     \end{equation}
    To show that such a choice of permutations exist, we will choose $\pi_1,\cdots,\pi_\ell$ independently and uniformly at random from the permutations of $W$ use Chernoff's inequality to bound the probability that a given copy $x$ of $K_{L-d}$ or $y$ of $T_{k-1}(L)$, fails to satisfy (4), and then take a union bound over the $2 \cdot L!$ choices of $x$ and $y$. 
    
    To spell out the details, let $\cE$ be the event that $F(x)=0$ for some $x\simeq K_{L-d}$ or $F(y)=1 $ for some $y\simeq T_{k-1}(L)$. Then

        $$\mathbb P(\cE)\leq \sum_{x\simeq K_{L-d}}\mathbb P\left(F(x)=0\right)+\sum_{y\simeq T_{k-1}(L)}\mathbb P\left(F(y)=1\right).$$
    Now, for any given $x\simeq K_{L-d}$ on $W$, we have
     $$\mathbb P(g(\pi_ix)=1)=a$$
     independently for every $i$, and therefore, by Chernoff
     $$\mathbb P \left(\sum_{i=1}^\ell g(\pi_ix)<\tau \right)\leq e^{-c\ell/m^2}$$
     and similarly, if $y\simeq T_{k-1}(L)$ on $W$ one has
     $$\mathbb P(g(\pi_iy)=1)=b$$
     independently for every $i$, and therefore, by Chernoff
     $$\mathbb P \left(\sum_{i=1}^\ell g(\pi_iy)\geq\tau \right)\leq e^{-c\ell/m^2}$$
     for some $c>0$. Therefore, recalling that $\ell =Cm^2L\log L$, a union bound shows that
     $$\mathbb P\left(\cE\right)\leq 2L!e^{-c\ell/m^2}=o(1)$$
     for $C>1/c$. Therefore, one may choose permutations $\pi_1,\ldots,\pi_\ell$ such that $F$ outputs $1$ on every copy of $K_{L-d}$ on $W$ and outputs $0$ on every copy of $T_{k-1}(L)$ on $W$. 
     
     Finally, observe that, by \Cref{thm:AKS}, there's a monotone circuit of size $O(\ell\log \ell)$ that computes $F$ given the $g(\pi_ix)$, and since by hypothesis we can compute each $g\pi_i$ with a monotone circuit of size $s$, it follows that
    $$\monsize F=O(\ell s+\ell\log \ell)$$
    as desired.
\end{proof}

\section{From cliques and Tur\'an graphs to matchings}\label{sec:reduction}

The objective of this section will be to show how one can go from a monotone circuit that can distinguish cliques and Tur\'an graphs to a monotone circuit of at most the same size, that can distinguish certain bipartite graphs with perfect matchings from those with no large matchings. Formally, we will prove the following lemma.

\begin{lemma}\label{lem:reduction}
    Let $k\leq L\leq 2(k-1)$, and set $r=L-k+1$ and $1\leq d< r$. If there is a monotone circuit on $L$-vertex graphs of size at most $s$ that can distinguish every $K_{L-d}$ from every $T_{k-1}(L)$, then there is a monotone circuit of size at most $s$ on the subgraphs of $K_{r,r}$ that can distinguish graphs with perfect matchings from those that don't have a matching of size $\min\{r-1,d+1\}$. 
\end{lemma}
By \Cref{thm:rao}, any circuit distinguishing bipartite graphs with large matchings from those with no large matchings will have to have large size. In particular, we will be able to prove the following corollary.
\begin{corollary}\label{coro:size_lower_bound} 
    There exists $c>0$ such that the following holds. Let $k\leq L\leq 2(k-1)$ and set $r=L-k+1$. For $1\leq d<r$ we have that any monotone circuit on $L$-vertex graphs that can distinguish copies of $K_{L-d}$ from copies of $T_{k-1}(L)$ has size at least $2^{cd/\sqrt r}$.
\end{corollary}
The proof of \Cref{lem:reduction} is straightforward. We take a circuit like in the hypothesis, and switch the $\con$ and $\dis$ gates to turn it into a circuit that distinguishes their complements. The complement of a copy of $T_{k-1}(L)$ will then be an $r$-matching, while the complement of a copy of $K_{L-d}$ will be the join of a $K_d$ and an independent set of size $L-d$. Restricting to the edges of some $K_{r,r}\subseteq K_L$ will give us the desired circuit

\begin{proof}[Proof of \Cref{lem:reduction}]
    Notice that any monotone circuit distinguishing copies of $K_{L-d}\subseteq K_L$ from copies of $T_{k-1}(L)$ can be turned into a monotone circuit of the same size that distinguishes their complements by simply switching all $\con$ gates to $\dis$ gates and vice-versa. Since $k\leq L\leq 2(k-1)$, a $T_{k-1}(L)$ consists of exactly $r$ parts of size $2$ and all other parts are singletons, therefore, its complement is exactly an $r$-matching. On the other hand, the complement of a $K_{L-d}\subseteq K_L$ is the join of a clique on $d$ vertices and an independent set of size $L-d$, we denote such a graph by $B_{d,L-d}$.

    Therefore, our task is to turn a circuit that can distinguish $r$-matchings from copies of $B_{d,L-d}$, into a circuit on $K_{r,r}$, of at most the same size, that can distinguish graphs with perfect matchings (i.e. $r$-matchings), from graphs with no matchings of size $q=\min\{r-1,d+1\}$. This can be done very directly. Arbitrarily choosing $2r\leq L$ vertices and partitioning them into two $r$-sets, while ignoring every edge that's not between them will be all that's needed. To see why this is true, notice that since our original circuit was monotone and it accepted $r$-matchings, this circuit will accept every subgraph of $K_{r,r}$ with a perfect matching. On the other hand, if a subgraph of $K_{r,r}$ does not have a $q$-matching, by K\H onig's theorem it must have a vertex cover of size $q-1\leq d$, and in particular, must be a subgraph of $B_{d,L-d}$. Since $B_{d,L-d}$ is rejected, our circuit will reject every subgraph of $K_{r,r}$ with no $q$-matching. This completes the proof.   
\end{proof}

\begin{proof}[Proof of \Cref{coro:size_lower_bound}]
    Let $f$ be a monotone circuit on $L$-vertex graphs that can distinguish copies of $K_{L-d}$ from copies of $T_{k-1}(L)$. By \Cref{lem:reduction}, there exists a monotone circuit $f'$ with $\monsize f'\leq\monsize f$ on the subgraphs of $K_{r,r}$ that can distinguish graphs with perfect matchings from those that don't have a matching of size $\min\{r - 1, d + 1\}$. Applying \Cref{thm:rao} with $N = 2r$ and $q = \min\{r - 1, d + 1\}$, we get
    $$\monsize f= 2^{\Omega(q/\sqrt r)}=2^{\Omega(d/\sqrt r)}$$
    as desired.

\end{proof}

\section{Proof of \Cref{thm:main}}\label{sec:main_proof}
We will show the following statement, which immediately implies \Cref{thm:main}
\begin{theorem}\label{thm:main_contrapositive}
    For every $a>0$ there is an $A>0$ such that if 
    $$\size_t\clique(n,\lfloor n/2\rfloor)\leq n^a$$ 
    then $t\geq\frac12\log n-\log\log n -A$.
\end{theorem}
\begin{proof}

Let $k=\lfloor n/2\rfloor$ and let $a>0$ be arbitrary. We will make choices for $m$ and $L_1,\ldots, L_m$ such that the hypotheses of \Cref{lem:size_bound} hold for $s=n^a$. To be precise, let $d:=Ca\sqrt n\log n$ for $C$ some large enough constant, and set
$$r_i=\frac k2+id\qquad \text{and}\qquad L_i=k+r_i-1.$$
Recall that this implies that the complement of the Tur\'an graph $T_{k-1}(L_i)$ is a matching of size $r_i$, as long as $r_i\leq k-1$. Let $m$ be maximal such that $r_m\leq k-1$. For this choice of $m$ we have
$$m=\Theta\left(\frac{\sqrt n}{a\log n}\right).$$
The following claim will feed into \Cref{lem:size_bound} to finish the proof

\begin{claim}\label{claim:false_positives}
    Let $g$ be a monotone circuit of size at most $n^a$, then one has
    \begin{equation}
        \E g(X_{L_i})-\E g(Y_{L_{i+1}})\leq m^{-1}\tag{4}
    \end{equation}
    for every $i\in \{0,\ldots, m-1\}$.
\end{claim}

\begin{proof}
    Since $L_i=L_{i+1}-d$, if (4) fails to hold then we are in the setting of \Cref{lem:approximate_to_exact} with $L=L_{i+1}$. Therefore, for $\ell=O(m^2L_{i+1}\log L_{i+1})=o(n^2)$, there must exist some monotone circuit $F$ on $K_{L_{i+1}}$ with
    $$\monsize F\leq \ell n^a+\ell\log\ell=o(n^{a+2})$$
    that distinguishes copies of $K_{L_{i+1}-d}$ from copies of $T_{k-1}(L_{i+1})$. But \Cref{coro:size_lower_bound} tells us such circuits must be large, explicitly, any such circuit must have size 
    $$\monsize F=\exp\left(\Omega\left(\frac{d}{\sqrt{r_{i+1}}}\right)\right).$$
    But $r_{i+1}=\Theta(n)$ and $d=Ca\sqrt n\log n$, and therefore, 
    $$\monsize F=2^{\Omega(Ca\log n)}\geq n^{a+2}$$
    for $C$ large enough. This is a contradiction.
\end{proof}

Now \Cref{lem:size_bound} gives
$$t\geq \log m-O(1)=\frac12\log n-\log\log n-O(\log a)$$
Taking $A$ large enough with respect to $a$ finishes the proof.
\end{proof}
\section{Concluding remarks}\label{sec:concluding_remarks}
By the result of Fischer \cite{F75} discussed above, our results are optimal up to a factor of $4$. Indeed, if our of $t$ could be improved to $\left\lceil\log\left(\binom n2+1\right)\right\rceil=2\log n+O(1)$ one would establish $P\neq NP$. Nevertheless, it seems the hard part of the task isn't in getting the correct constant, but in dealing with the lower-order terms. Indeed, it was shown recently by Karde\c s \cite{K26}, building on the work of Jukna \cite{J04}, that it is not enough in general to get the correct constant. Explicitly, there exist monotone functions $f:\{0,1\}^n\to\{0,1\}$ that admit polynomially sized circuits, and in particular, by Fischer, admit polynomially sized circuits with at most $\left\lceil\log\left(n+1\right)\right\rceil$ negations but nevertheless, 
$$\size_t f=n^{\omega(1)}\qquad\text{for }\qquad t\leq \log n-O(\log\log n).$$

\section*{Acknowledgements}
The author would like to thank Gabriel Dahia for suggesting the problem to him, and Rob Morris for helpful comments on an earlier draft of this paper.

\end{document}